\documentclass[a4paper, 12pt]{article}

\usepackage[english]{babel}
\usepackage{cite}
\usepackage{a4wide}
\usepackage{amsmath}
\usepackage[charter,cal=cmcal,greekuppercase=italicized]{mathdesign}
\usepackage{dsfont}
\usepackage[T1]{fontenc}
\usepackage{slashed}
\usepackage{setspace}
\usepackage{graphicx}
\usepackage{xcolor}
\usepackage{soul}
\usepackage[final,          
            colorlinks,     
            linkcolor=purple, 
            citecolor=teal, 
            urlcolor=blue  
            ]{hyperref}

\makeatletter
\DeclareRobustCommand*{\bfseries}{%
  \not@math@alphabet\bfseries\mathbf
  \fontseries\bfdefault\selectfont
  \boldmath
}
\makeatother

\newcommand{\im}{\ensuremath{\mathrm{i}\,}}

\newcommand{\Tr}{\operatorname{Tr}}
\newcommand{\diag}{\operatorname{diag}}

\newcommand{\Lag}{\ensuremath{\mathcal{L}}}

\newcommand{\U}{\ensuremath{\mathrm{U}}}

\newcommand{\CP}{\ensuremath{CP}}
\newcommand{\eV}{\ensuremath{\mathrm{eV}}}

\newcommand{\GeV}{\ensuremath{\mathrm{GeV}}}
\newcommand{\TeV}{\ensuremath{\mathrm{TeV}}}

\newcommand{\hc}{\ensuremath{\text{h.\,c.\ }}}

\renewcommand{\Re}{\ensuremath{\operatorname{Re}}}

\newcommand{\Mat}[1]{\ensuremath{\boldsymbol{#1}}}

\newcommand{\ie}{i.\,e.~}

\newcommand{\eg}{e.\,g.~}

\usepackage[font=small]{caption}
\usepackage{graphicx}

\begin{document}

\onehalfspacing

\begin{titlepage}

\vspace*{-15mm}
\begin{flushright}
DESY 17-221 \\
TTP 17-051
\end{flushright}
\vspace*{0.7cm}

\begin{center} {
\bfseries\LARGE
Texture zeros and hierarchical masses \\[2mm]
from flavour (mis)alignment
}\\[8mm]
W.~G.~Hollik $^{a,}$
\footnote{\texttt{w.hollik@desy.de}}\quad
and\quad
U.~J.~Saldana-Salazar $\,^{b,}$
\footnote{\texttt{ulises.saldana-salazar@partner.kit.edu}}
\\[1mm]
\end{center}
\vspace*{0.50cm}
\centerline{$^a$ \itshape
Deutsches Elektronen-Synchrotron (DESY),}
\centerline{\itshape
Notkestra{\ss}e 85, D-22607 Hamburg, Germany.}
\centerline{$^b$ \itshape
Institut f\"ur Theoretische Teilchenphysik (TTP),
Karlsruhe Institute of Technology,}
\centerline{\itshape Engesserstra{\ss}e 7, D-76131 Karlsruhe, Germany.}
\vspace*{1.20cm}

\begin{abstract}
  We introduce an unconventional interpretation of the fermion mass
  matrix elements. As the full rotational freedom of the gauge-kinetic
  terms renders a set of infinite bases called weak bases,
  basis-dependent structures as mass matrices are unphysical. Matrix
  invariants, on the other hand, provide a set of basis-independent
  objects which are of more relevance. We employ one of these invariants
  to give a new parametrisation of the mass matrices. By virtue of it,
  one gains control over its implicit implications on several mass
  matrix structures. The key element is the trace invariant which
  resembles the equation of a hypersphere with a radius equal to the
  Frobenius norm of the mass matrix. With the concepts of alignment or
  misalignment we can identify texture zeros with certain alignments
  whereas Froggatt--Nielsen structures in the matrix elements are
  governed by misalignment. This method allows further insights of
  traditional approaches to the underlying flavour geometry.
\end{abstract}

\end{titlepage}

\setcounter{footnote}{0}

\section{Introduction}

After different trials to understand the various unsolved aspects of
fermion masses and mixing, the so called \emph{flavour puzzle} still
lacks for a satisfactory explanation. In spite of this, some hints could
already be pointing out for a theory of flavour, see for example
Refs.~\cite{Hollik:2014jda, Saldana-Salazar:2015raa, Diaz-Cruz:2016pmm,
  Saldana-Salazar:2016pms, Saldana-Salazar:2016hxb}. The common
approaches have mainly concerned on introducing zeros (texture zeros) in
the mass matrices in order to reduce the number of
parameters~\cite{Fritzsch:1977vd, Weinberg:1977hb, Branco:1988iq,
  Ramond:1993kv, Branco:1994jx, Branco:1999nb, Branco:2010tx,
  Emmanuel-Costa:2016gdp}, the use of flavour symmetries which at the
same time can justify some of the aforementioned
zeros~\cite{Ishimori:2010au}, the use of hierarchical fermion masses to
unveil the structure in fermion mixing~\cite{Rasin:1998je,
  Hollik:2014jda}, the Froggatt--Nielsen
mechanism~\cite{Froggatt:1978nt} or extra dimensions to produce
hierarchical fermion masses and mixing angles~\cite{ArkaniHamed:1999dc},
among others.

The main puzzle arises from the \emph{complete arbitrariness} in which
the mass matrices appear in the Standard Model (SM), proportional to the
Yukawa couplings of fermions to the Higgs field, such that after
electroweak symmetry breaking, a generic fermion mass matrix is given by
\begin{equation}\label{eq:massmatrix}
        \Mat{M} = \frac{v}{\sqrt{2}}
\begin{pmatrix}
        |y_{11}| e^{i\delta_{11}} & |y_{12}| e^{i\delta_{12}} & |y_{13}| e^{i\delta_{13}}  \\
        |y_{21}| e^{i\delta_{21}} & |y_{22}| e^{i\delta_{22}} & |y_{23}| e^{i\delta_{23}}  \\
        |y_{31}| e^{i\delta_{31}} & |y_{32}| e^{i\delta_{32}} & |y_{33}| e^{i\delta_{33}}
\end{pmatrix},
\end{equation}
with $v = 246\text{ GeV}$ the Higgs vacuum expectation value. There are
in general much more parameters allowed than physical. Moreover, the
question why there are three generations, so why are they \(3 \times 3\)
matrices stays unclear. We do not intend to resolve this open question
here but rather like to scrutinise the underlying arbitrariness. A new
level of understanding may be gained by a study of the generic
properties of these mass matrices and identification which or how many
of the available parameters can be physical at the end. Later, one may
find a fundamental reason behind its construction. Regarding this
two-level approach, in this letter, we provide a way to dissolve the
initial arbitrariness and understand some of the phenomenological
observations that have already been made. The second part lies beyond
the scope of our present work.

In the limit of massless fermions, \eg vanishing Yukawa couplings, the
matter sector of the Standard Model reveals a very large accidental
symmetry. This symmetry allows for some arbitrariness in the choice of
a weak basis.\footnote{A weak basis is a particular choice of
  \(\U(3)\) transformations which leave the neutral and charged
  current interactions invariant.}  The largest flavour symmetry is
given by the following global symmetries on the fermion fields:
\begin{equation}
  \mathcal{G}_F \supset \U(3)_L^F \times \U(3)_R^a\times \U(3)_R^b ,
\end{equation}
which holds for both quarks and leptons, where \(F = Q, \ell\) stands
for the left-handed doublet fields and $a=u,\nu$ and $b = d, e$ for the
right-handed singlets if we add 3 right-handed neutrinos to the Standard
Model to be symmetric in the quark and lepton sector.\footnote{In
  general, models for neutrino masses involve a much broader range of
  possibilities. For our study, the explicit UV complete theory of
  neutrino masses does not play a role and we can even work with the
  field content of the pure SM only (no right-handed neutrinos and only
  an effective mass operators for the light neutrinos).} The mass
matrices \(\Mat{M}_a\) and \(\Mat{M}_b\) are modified by these weak
basis transformations,
\begin{equation} \label{eq:genSVD}
  \Mat{M}'_a = \Mat{L}_Q \Mat{M}_a \Mat{R}_a^\dag
  \qquad \text{and} \qquad
  \Mat{M}'_b = \Mat{L}_Q \Mat{M}_b \Mat{R}_b^\dag.
\end{equation}
 where left- and
right-handed fields are transformed independently
\begin{subequations}
\begin{align}
\psi^F_L & \to \Mat{L}_F , \\
\psi^a_R & \to \Mat{R}_a, \\
\psi^b_R & \to \Mat{R}_b,
\end{align}
\end{subequations}
with \(\Mat{X}_y \in \U(3)_X^y\) unitary transformations, meaning
\(\Mat{X}_y^\dag \Mat{X}_y = \Mat{X}_y \Mat{X}_y^\dag = \Mat{1}\).

Basically, this ambiguity reveals \((3\times 9) = 27\) free parameters
which have to be balanced with \((9\times 2 \times 2) = 36\) arbitrary
parameters in the mass matrices like Eq.~\eqref{eq:massmatrix}. In
addition, there is a freedom of a global rephasing in each fermion
sector, known as global baryon or lepton number which remains after
introducing the masses. Thus, the number of physical
parameters\footnote{Unphysical is the full rotational freedom of the
  gauge-kinetic terms.} apparently is given by \(36 - 27 + 1 = 10\)
which decomposes to the six masses, three mixing angles and one complex
phase. In the case of \eg light Majorana neutrinos, their mass matrix is
constrained to be complex, but symmetric, so the counting is slightly
different, especially because no \(\U(3)_R^\nu\) freedom exists. We then
have \(2\times 9\) arbitrary parameters from the complex \(3\times 3\)
charged lepton masses and \(2\times 6 = 12\) parameters from the complex
symmetric neutrino Majorana mass, see also
Section~\ref{sec:majorana}. In total, we are left with \(30 - 18 = 12\)
physical parameters: compared to the pure Dirac case there are two more
complex phases, the well-known Majorana phases.

In the course of this letter, we present a novel route on how to relate
the initially free parameters of the mass matrices with the weak basis
transformations and define a new \emph{interpretation} for the
individual mass matrix elements on a geometrical argument. By
geometrical reasoning (as \eg alignment/misalignment), we can dissolve
the arbitrariness within a weak basis and give a way to study underlying
flavour patterns through a systematical procedure. While there exists
already an exhaustive literature on the problem how weak basis
transformations affects flavour structures and texture zeros in a
general way, see \eg Refs.~\cite{Branco:1988iq, Branco:1994jx,
  Branco:1999nb, Branco:2010tx, Emmanuel-Costa:2016gdp}, our geometrical
approach differs from them in its easiness and originality.

This letter is organised as follows: in Section~\ref{sec:reparam}, we
propose a new spherical parametrisation for the magnitude of the mass
matrix elements following from the matrix invariants. In
Section~\ref{sec:interp}, we relate the angles of the spherical mass
matrix to the physical angles and discuss an explicit two-family
description in Section~\ref{sec:2flavour}.  We examine the nature of
texture zeros in Section~\ref{sec:zeros} and in
Section~\ref{sec:majorana} we explore similar considerations for the
case of Majorana neutrinos. The description of large fermion mass
hierarchies by small angles can be found in
Section~\ref{sec:smallangles} relating to Froggatt--Nielsen-like
models. Finally, in Section~\ref{sec:concl} we conclude.

\section{The spherical mass matrix interpretation}
\label{sec:reparam}

Let $\Mat{M}$ be a generic $3\times 3$ complex mass matrix,
\begin{equation}
  \Mat{M} = \begin{pmatrix}
    m_{11} & m_{12} & m_{13} \\
    m_{21} & m_{22} & m_{23} \\
    m_{31} & m_{32} & m_{33}
  \end{pmatrix}.
\end{equation}
Its Singular Value Decomposition (SVD) is given as
\begin{equation}
        \Mat{M} = \sum_{j=1}^3 \ell_j m_j r_j^\dag
\end{equation}
where $\ell_j$ and $r_j$ are the singular vectors corresponding to the
\(j\)-th singular value (mass) $m_j$. They set up the left and right
unitary transformations \(\Mat{L}\) and \(\Mat{R}\) of
Eq.~\eqref{eq:genSVD}, which diagonalise the two hermitian products of
\(\Mat{M}\): \(\Mat{L}^\dag \Mat{M} \Mat{M}^\dag \Mat{L} =
\diag(m_1^2,m_2^2,m_3^2)\) and \(\Mat{R}^\dag \Mat{M}^\dag \Mat{M}
\Mat{R} = \diag(m_1^2,m_2^2,m_3^2)\), respectively.

A complex \(3 \times 3\) matrix has three invariants that do not change
under the left and right unitary transformations:
\begin{align}
  \xi = \frac{1}{2} \left[ \Tr\left[ \Mat{M} \Mat{M}^\dag \right]^2
    - \Tr\left[ \left( \Mat{M} \Mat{M}^\dag \right)^2 \right] \right]
  &=  m_1^2 m_2^2 + m_2^2 m_3^2 + m_1^2 m_3^2, \\
  D = \det\left[ \Mat{M} \Mat{M}^\dag \right] &= m_1^2 m_2^2 m_3^2, \\
  R^2 = \Tr\left[ \Mat{M} \Mat{M}^\dag \right] &= m_1^2 + m_2^2 + m_3 ^2,
\end{align}
which can be expressed in terms of the singular values or
masses. Conversely, this same set can be written using the mass matrix
elements,
\begin{align}
  \xi &= x_1 x_2 + x_1 x_3 + x_2 x_3 - (|y_1|^2 + |y_2|^2 + |y_3|^2), \\
  D &=  x_1 x_2 x_3 - x_1 |y_3|^2 - x_2 |y_2|^2 - x_3 |y_1|^2
  + 2\Re(y_1 y_2^* y_3), \\
  R^2 &= x_1 + x_2 + x_3,
\end{align}
where we have abbreviated
\begin{subequations} \label{eq:xs-and-ys}
\begin{align}
 x_1 &=  |m_{11}|^2 + |m_{12}|^2 + |m_{13}|^2,  \\
 x_2 &=  |m_{21}|^2 + |m_{22}|^2 + |m_{23}|^2,  \\
 x_3 &=  |m_{31}|^2 + |m_{32}|^2 + |m_{33}|^2, \\
 y_1 &= m_{11} m_{21}^* + m_{12} m_{22}^* + m_{13} m_{23}^*, \\
 y_2 &=  m_{11} m_{31}^* + m_{12} m_{32}^* + m_{13} m_{33}^*, \\
 y_3 &= m_{21} m_{31}^* + m_{22} m_{32}^* + m_{23} m_{33}^*.
\end{align}
\end{subequations}
Of course, all these equations are well-known facts and these relations
already have been exploited in the flavour physics context, see \eg
Ref.~\cite{Couture:2009it,Branco:2011aa}. Nevertheless, we want to state
a very pictorial \emph{interpretation}, which can be shown to be a
powerful parametrisation of the mass matrix arbitrariness. In this
interpretation the trace invariant suggests a parametrisation of the
matrix elements describing the surface of a hypersphere. As can be
easily seen, the trace of the hermitian product is given by the sum of
squared matrix elements which also defines the Frobenius norm
\(||{\Mat{M}}||_F\). Thus, we have the relation
\begin{equation} \label{eq:FrobNorm}
  R^2 = \Tr\left[ \Mat{M} \Mat{M}^\dag \right]
  = ||{\Mat{M}}||_F^2
  = \sum_{i,j} |m_{ij}|^2.
\end{equation}
This is the equation of a hypersphere in \(n^2\) dimensions, for \(i,j =
1, \ldots n\) and \(n=2,3\) for most of our purposes. It suggests a very
elegant way of parametrising the individual matrix elements in terms of
spherical coordinates.

In the following, we define a slightly different notion of flavour space
than what is usually understood. Mass terms are usually written in terms
of Lorentz-invariants and are explicitly flavour dependent. If we wished
to introduce flavour invariance we would find that it requires a more
careful treatment. The notion of a flavour symmetry or a democratic
approach as the one proposed in Ref.~\cite{Saldana-Salazar:2015raa} are
part of some of the trials to extend the flavour invariance of the
kinetic terms to the Yukawa sector.

Let us already put our personal bias in the choice of coordinate system.
The final values, however, do not depend explicitly on that choice as
always a certain transformation can be found that redefines the
axes.\footnote{This freedom can be characterised by the independent
  permutation of columns and rows $S_{3L}\times S_{3R}$, where $S_{3}$
  is the group of permutations of three identical objects.}  For the
hypersphere equation \eqref{eq:FrobNorm}, the complex nature of the
matrix elements plays no role, so for the following we consider a
\emph{real} \(3\times 3\) matrix
\begin{equation}
\widetilde{\Mat{M}} = \begin{pmatrix}
 \widetilde{m}_{11} & \widetilde{m}_{12} & \widetilde{m}_{13} \\
 \widetilde{m}_{21} & \widetilde{m}_{22} & \widetilde{m}_{23} \\
 \widetilde{m}_{31} & \widetilde{m}_{32} & \widetilde{m}_{33}
\end{pmatrix},
\end{equation}
with
\begin{subequations} \label{eq:sph-matrix-elems}
\begin{align}
\widetilde{m}_{11} &= R \sin\chi \sin\phi_1 \sin\phi_2 \sin\phi_3 \sin\phi_4
\sin\phi_5 \sin\phi_6 \sin\phi_7, \\
\widetilde{m}_{12} &= R \sin\chi \sin\phi_1 \sin\phi_2 \sin\phi_3 \sin\phi_4
\sin\phi_5 \sin\phi_6 \cos\phi_7, \\
\widetilde{m}_{13} &= R \sin\chi \sin\phi_1 \sin\phi_2 \sin\phi_3 \sin\phi_4
\sin\phi_5 \cos\phi_6, \\
\widetilde{m}_{21} &= R \sin\chi \sin\phi_1 \sin\phi_2 \sin\phi_3 \sin\phi_4
\cos\phi_5, \\
\widetilde{m}_{22} &= R \sin\chi \sin\phi_1 \sin\phi_2 \sin\phi_3 \cos\phi_4,
\\
\widetilde{m}_{23} &= R \sin\chi \sin\phi_1 \sin\phi_2 \cos\phi_3, \\
\widetilde{m}_{31} &= R \sin\chi \sin\phi_1 \cos\phi_2, \\
\widetilde{m}_{32} &= R \sin\chi \cos\phi_1, \\
\widetilde{m}_{33} &= R \cos\chi.
\end{align}
\end{subequations}
The angles are \(\phi_i \in [0, 2\pi)\), \(i=1,\ldots,7\), and \(\chi
\in [0, \pi]\). The mass matrix is then written as,
\begin{equation}
\widetilde{\Mat{M}} = R \begin{pmatrix}
 \sin\chi \left(\prod_{i=1}^6 \sin\phi_i \right) \sin\phi_7 &
 \sin\chi \left(\prod_{i=1}^6 \sin\phi_i \right) \cos\phi_7 &
 \sin\chi \left(\prod_{i=1}^5 \sin\phi_i \right) \cos\phi_6 \\
 \sin\chi \left(\prod_{i=1}^4 \sin\phi_i \right) \cos\phi_5 &
 \sin\chi \left(\prod_{i=1}^3 \sin\phi_i \right) \cos\phi_4 &
 \sin\chi \left(\prod_{i=1}^2 \sin\phi_i \right) \cos\phi_3 \\
 \sin\chi \sin\phi_1 \cos\phi_2 &
 \sin\chi \cos\phi_1 &
 \cos\chi
\end{pmatrix}.
\end{equation}
Although it does not look very advantageous to express the mass matrix
elements like this, we can immediately draw some useful applications
out. First, we see directly how the matrix elements can be interrelated:
an adjustment in one element also affects the others unless it means
exact alignment in one angle or only a small misalignment. Second, we
can with a certain choice of angles immediately produce ``texture
zeros'': null mass matrix elements at distinct positions. For example, a
vanishing \(m_{11}\) then could be obtained by setting \(\phi_7 = 0\)
without severely influencing any other matrix element (notice that
\(\cos\phi_7 = 1\) in \(m_{12}\) and the angle appears nowhere
else). Similarly, for \(m_{13} = 0\) one chooses \(\phi_6 =
\frac{\pi}{2}\), and so on. Third, we discover that
Froggatt--Nielsen-like patterns can easily be produced for small angles,
see Section~\ref{sec:smallangles}: misalignment instead of alignment.
We are going to give a more physical connection to the observable and
well-known flavour angles in Section \ref{sec:interp}.

It is easy to relate the mass matrix entries in this interpretation as a
9-dimensional vector
\[
\overrightarrow{\mathfrak{m}} = (
\widetilde{m}_{11}, \widetilde{m}_{12}, \widetilde{m}_{13},
\widetilde{m}_{21}, \widetilde{m}_{22}, \widetilde{m}_{23},
\widetilde{m}_{31}, \widetilde{m}_{32}, \widetilde{m}_{33} )^T
\]
to some flavour space, where we define the axes accordingly:
\begin{equation}
-\mathcal{L} = \sum_{i,j=1}^3 \overline \psi_{L,i} \widetilde m_{ij}
\psi_{R,j} \equiv \sum_{i,j=1}^3 \widetilde m_{ij} \hat x_{ij},
\end{equation}
with \(\hat x_{ij}\) a unit vector in the \(i\)-\(j\) direction, where
the first index refers to the left-handed fermions and the second one to
the right-handed. Surely, the individual \(\hat x_{ij}\)-directions
cannot be treated independently as they are the outer product of some
flavour vectors and calculus rules for outer products
apply. Nevertheless, we consider the vectors \(\hat x_{ij}\) as basis of
the 9 dimensional vector space spanned by the mass matrix elements
describing the surface of a hypersphere. The apparent redundancy gets
reduced later on.

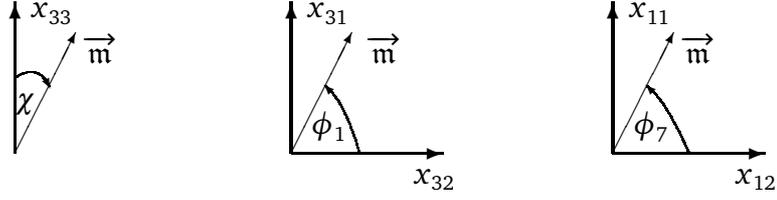
\begin{figure}
\centering
\setlength{\unitlength}{1cm}
\begin{picture}(2,2)
\put(0.5,0.4){\vector(1,2){0.8}}
\put(1.4,1.6){\(\overrightarrow{\mathfrak{m}}\)}
\put(0.7,2.2){\(x_{33}\)}
\qbezier(0.5,1.4)(0.8,1.6)(0.95,1.29)
\put(0.95,1.29){\vector(1,-2){0.02}}
\put(0.5,1){\(\chi\)}
\thicklines
\put(0.5,0.4){\vector(0,1){2}}
\end{picture}
\qquad\qquad
\begin{picture}(2.5,2.5)
\put(2,0){\(x_{32}\)}
\put(0.6,2.2){\(x_{31}\)}
\put(0.4,0.4){\vector(1,2){0.8}}
\put(1.4,1.6){\(\overrightarrow{\mathfrak{m}}\)}
\qbezier(1.3,0.4)(1.1,1.1)(0.85,1.29)
\put(0.85,1.29){\vector(-1,1){0.02}}
\put(0.66,0.66){\(\phi_1\)}
\thicklines
\put(0.4,0.4){\vector(0,1){2}}
\put(0.4,0.4){\vector(1,0){2}}
\end{picture}
\qquad\qquad
\begin{picture}(2.5,2.5)
\put(2,0){\(x_{12}\)}
\put(0.6,2.2){\(x_{11}\)}
\put(0.4,0.4){\vector(1,2){0.8}}
\put(1.3,1.6){\(\overrightarrow{\mathfrak{m}}\)}
\qbezier(1.4,0.4)(1.1,1.1)(0.85,1.29)
\put(0.85,1.29){\vector(-1,1){0.02}}
\put(0.66,0.66){\(\phi_7\)}
\thicklines
\put(0.4,0.4){\vector(0,1){2}}
\put(0.4,0.4){\vector(1,0){2}}
\end{picture}
\caption{Visualization of the angles \(\chi\), \(\phi_1\), and
  \(\phi_7\). The other \(\phi_i\) follow analogously; the coordinate
  \(x_{ij}\) represents the axis relating \(i\)-th and \(j\)-th
  generation \(\sim \bar\psi_{L,i} \psi_{R,j}\).}
\label{fig:alignment}
\end{figure}

In this interpretation, it can be easily seen that the angle \(\chi\)
represents the deviation of the mass vector
\(\overrightarrow{\mathfrak{m}}\) from the 3-3 axis (\(\chi = 0\) means
full alignment with the third generation of left- and right-handed
fields\footnote{It is interesting to notice how this is approximately
  true for the known values of the charged fermion masses.}). The other
angles represent the relative orientation with respect to two axes, so
\(\phi_1\) interpolates between the 3-2 and the 3-1 axis and \(\phi_2\)
between 3-1 and 2-3 and so forth, see Fig.~\ref{fig:alignment}. Notice
that in our specific parametrisation from above, the last angle
\(\phi_7\) has the axis flipped with respect to the usual convention
(\ie\ in three dimensions) and \(\phi_7 = 0\) means alignment with the
1-2 axis rather than 1-1, which is very useful for the application in
flavour physics.

\section{Relating mass matrix elements to physical angles}
\label{sec:interp}

We show briefly in the following how the eight angles in the spherical
mass matrix interpretation can be related to the physical angles in the
mixing matrices and masses. The Frobenius norm of a general complex and
rectangular \(m \times n\) matrix \(\Mat{A}\) is given by the square
root of the sum of its matrix elements \(a_{ij}\) squared,
\begin{equation}
  ||\Mat{A}||_{F} = \sqrt{\Tr\left( \Mat{A} \Mat{A}^\dag \right)}
  = \sqrt{\sum_{i=1}^m \sum_{j=1}^n |a_{ij}|^2}.
\end{equation}
In return, this relation may be seen as an hypersphere equation in \(m
\times n\) dimensions with the Frobenius norm as radius of the
sphere. The corresponding spherical coordinates require \((m\times n
-1)\) angles and one radius.

On the other hand, this complex matrix has a number of $q$ non-zero and
positive singular values, $\sigma_i>0$. This defines its rank to be
$q$. The Frobenius norm can also be expressed in terms of the singular
values as
\begin{equation}\label{eq:FrobNorm2}
  ||\Mat{A}||_{F} = \sqrt{\sum_{i=1}^q \sigma_i^2},
\end{equation}
and similarly, this characterises the surface of a $q$-dimensional
hypersphere.

For the following, we restrict ourselves to the flavour-physical case of
square matrices, in particular with dimension three. We work on the
surface of unit sphere, where the radius is an overall scaling factor
and can be factored out by normalizing the matrix to its Frobenius norm
\begin{equation} \label{eq:normalised}
  \bar{\Mat{A}} =
  \frac{\Mat{A}}{||\Mat{A}||_F}.
\end{equation}
For the normalised singular values, we define
\begin{equation} \label{eq:sing-val-sph}
  \bar \sigma_1 =   \sin \alpha \sin \beta, \qquad
  \bar \sigma_2 =   \sin \alpha \cos \beta, \qquad
  \bar \sigma_3 =   \cos \alpha,
\end{equation}
with \(\alpha, \beta \in [0, \tfrac{\pi}{2}]\) for all
\(\bar\sigma_i>0\). The three matrix invariants expressed through
Eqs.~\eqref{eq:sing-val-sph} are then
\begin{align}
  \bar{R}^2 = \Tr \left( \bar{\Mat{A}}  \bar{\Mat{A}}^\dag \right) &= 1, \\
  \bar D = \det \left( \bar{\Mat{A}}  \bar{\Mat{A}}^\dag \right) &=
    \sin^4 \alpha \sin^2 \beta \cos^2 \alpha \cos^2 \beta, \\
    \bar \xi = \tfrac{1}{2} \left[ \Tr\left[ \bar{\Mat{A}}
        \bar{\Mat{A}}^\dag \right]^2 - \Tr\left[ \left( \bar{\Mat{A}}
          \bar{\Mat{A}}^\dag \right)^2 \right] \right] &=
      \sin^2 \alpha \left( \sin^2 \alpha \sin^2 \beta \cos^2 \beta
        + \cos^2 \alpha \right).
\end{align}

Eq.~\eqref{eq:sing-val-sph} shows that two angles are enough to
describe the normalised singular values spectra, which is equivalent
to the fact that only two independent mass ratios are relevant. This
can be trivially extended to the $n$ family case.

The next step is to reconsider the hypersphere made out of the matrix
elements which carries more information than the singular value
spectrum. In this case, a nine dimensional hypersphere requires a set of
eight angles as written in Eq.~\eqref{eq:sph-matrix-elems}. These eight
angles are to be related with the two ``angles'' describing the span of
the singular values and furthermore \(2 \times 3\) from the left and
right unitary rotations. The two angles most tightly related to the
singular values can be read from comparison with
Eq.~\eqref{eq:sph-matrix-elems} and we find $\chi$ and $\phi_3$ to be
important here. The other six angles, however, have to be related via
the usual SVD
\begin{equation}\label{eq:svd}
  \Mat{M}_f = \Mat{L}_f^\dag \Mat{\Sigma}_f \Mat{R}_f,
\end{equation}
where we have three mixing angles in \(\Mat{L}_f\) and \(\Mat{R}_f\)
each. Furthermore, the unitary transformations acting on the
left-handed fields are physical in the sense that their combined
product
\begin{equation}
  \Mat{V} = \Mat{L}_a \Mat{L}_b^\dag,
\end{equation}
describes the mixing matrix of the charged current interaction and thus
the angles of \(\Mat{V}\) are the observable quantities. The right-handed
rotations disappear from phenomenology.

The SVD is independent of the normalisation factor and is given in an
explicit form with the singular values of Eq.~\eqref{eq:sing-val-sph}
\begin{equation}\label{eq:svd-angle}
  \bar{\Mat{M}} \equiv \frac{\Mat{M}}{||\Mat{M}||_F}=
  \Mat{L}^\dag \begin{pmatrix}
    \sin \alpha \sin \beta  & 0 & 0\\
    0 & \sin \alpha \cos \beta & 0 \\
    0 & 0 & \cos \alpha
  \end{pmatrix} \Mat{R} .
\end{equation}
The unitary transformations $\Mat{L}$ and $\Mat{R}$ can be parametrised
by three angles and six complex phases each. Some of the phases are
redundant and can be absorbed in the fermion fields, so let us for
simplicity first study the rotation matrices as real matrices.  The
right hand side of Eq.~\eqref{eq:svd-angle} embraces eight independent
angles: two from the singular values and three coming from each unitary
transformation, the same amount as in \(\bar{\Mat{M}}\).\footnote{This
  observation is rather trivial, since the number of independent
  parameters has to be balanced on the two sides, and for the SVD an
  overall factor plays no role.} The right transformations \(\Mat{R}\),
however, are unphysical in the sense that they drop out from physical
observables and only the left rotations \(\Mat{L}\) play a
role. Furthermore, whenever the same left transformation \(\Mat{L}\) is
used in both mass matrices, the charged current remains invariant, so
this adds three more unobservable angles.  Hence, from the right
transformations, there are three unphysical angles for up- and down-type
fermions each, whereas from the left ones, three more are included to
the sum, reaching a total of nine unphysical angles; this freedom can be
used \eg to remove mass matrix elements, \ie introduce ``texture
zeros''. The singular values in the reduced form lack one more parameter
each, which is the Frobenius norm and sets the scale of the largest
mass.

It is then a simple task to determine the ``angles'' \(\alpha\) and
\(\beta\) as functions of the (normalised) singular values. With the
definition of Eq.~\eqref{eq:sing-val-sph}, the \(\bar\sigma_i\) are the
singular values of the matrix \(\bar{\Mat{M}}\) and one easily finds
\(\tan\beta = \bar\sigma_1 / \bar\sigma_2\) and correspondingly
\(\sin\beta \tan\alpha = \bar\sigma_1 / \bar\sigma_3\) for the ratios of
first to second and third generation masses. So we have the
identities,\footnote{We employ \(\sin(\arctan x) =
  \frac{x}{\sqrt{1+x^2}}.\)}
\begin{subequations}
\begin{align}
  \sin\beta &= \sqrt{\frac{\bar\sigma_1^2}{\bar\sigma_1^2 + \bar\sigma_2^2}}, \\
  \sin\alpha &= \sqrt{\frac{\bar\sigma_1^2 +
      \bar\sigma_2^2}{\bar\sigma_1^2 + \bar\sigma_2^2 +
      \bar\sigma_3^2}}.
\end{align}
\end{subequations}

\section{The two-flavour philosophy}
\label{sec:2flavour}
Although two-flavour scenarios mostly lack the complexity of the
``true'' three-family construction, it is very helpful to see what is
going on and provide a gateway to further complications.

Let us consider an arbitrary \(2 \times 2\) mass matrix,
\begin{equation}
\Mat{m} = \begin{pmatrix} m_{11} & m_{12} \\ m_{21} & m_{22} \end{pmatrix},
\end{equation}
with real matrix elements \(m_{ij}\). A singular value decomposition
of this matrix is given by \(\Mat{m} = \Mat{L}^\dag \Mat{\Sigma}
\Mat{R}\) with \(\U(2)\)-matrices \(\Mat{L}\) and \(\Mat{R}\) and the
diagonal matrix of singular values \(\Mat{\Sigma} = \diag(\sigma_1,
\sigma_2) \) with \(\sigma_2 \geqslant \sigma_1 > 0\). The matrix
invariants relate the (somewhat arbitrary) entries of \(\Mat{m}\) with
the singular values, so from the trace,
\begin{equation}
\Tr\left[ \Mat{m} \Mat{m}^\dag \right] = |m_{11}|^2 + |m_{12}|^2 + |m_{21}|^2 +
  |m_{22}|^2 = \sigma_1^2 + \sigma_2^2 = \Tr\left[ \Mat{\Sigma}^2
  \right] \equiv r^2.
\end{equation}
This equation constrains the matrix elements to the surface of a
four-dimensional sphere and also correlates the two singular values
with a circle, \(\sigma_1 = r \sin\zeta\) and \(\sigma_2 = r
\cos\zeta\) with \(\zeta \in [0, \frac{\pi}{2}]\) to avoid any
negative \(\sigma_k\). Consequently, we can write
\begin{equation} \label{eq:2fam-wb-angles}
\begin{aligned}
\Mat{m}
&= r \begin{pmatrix}
  \cos\theta_L & - \sin\theta_L \\
  \sin\theta_L & \cos\theta_L
\end{pmatrix}
\begin{pmatrix} \sin\zeta & 0 \\  0 & \cos\zeta \end{pmatrix}
\begin{pmatrix}
  \cos\theta_R & \sin\theta_R \\
  - \sin\theta_R & \cos\theta_R
\end{pmatrix} \\
&= r \begin{pmatrix}
\sin\zeta \cos\theta_L \cos\theta_R + \cos\zeta \sin\theta_L \sin\theta_R &
\sin\zeta \cos\theta_L \cos\theta_R - \cos\zeta \sin\theta_L \cos\theta_R \\
\sin\zeta \sin\theta_L \cos\theta_R - \cos\zeta \cos\theta_L \sin\theta_R &
\cos\zeta \cos\theta_L \cos\theta_R + \sin\zeta \sin\theta_L \sin\theta_R
\end{pmatrix}.
\end{aligned}
\end{equation}
It is very intriguing to also look at the left-symmetric product in this
way and discuss its relation to the choice of a weak basis. We have
\begin{equation}
\Mat{m} \Mat{m}^\dag = \frac{r^2}{2}
\begin{pmatrix}
1- \cos(2\zeta)\cos(2\theta_L) & - \cos(2\zeta)\sin(2\theta_L) \\
-\cos(2\zeta)\sin(2\theta_L) & 1 + \cos(2\zeta)\cos(2\theta_L)
\end{pmatrix},
\end{equation}
what trivially tells us, that \(\theta_L = 0\) is the basis in which
\(\Mat{m} \Mat{m}^\dag\) is diagonal and \(\Deltaup \sigma^2 =
\sigma_2^2 - \sigma_1^2 = \cos(2\zeta)\).

On the other hand, we can also use the spherical mass matrix
interpretation to find
\begin{equation}
  \Mat{m} = r
\begin{pmatrix}
  \sin\chi \sin\phi_1 \sin\phi_2 & \sin\chi \sin\phi_1 \cos\phi_2 \\
  \sin\chi \cos\phi_1 & \cos\chi
\end{pmatrix}.
\end{equation}
It is not a straightforward task to build a direct connection between
these angles and those appearing in
Eq. \eqref{eq:2fam-wb-angles}.\footnote{A similar structure, however,
  can appear if instead of rotating flavour space one \emph{shears}
  it. So, \eg one finds
\[\begin{aligned}
\Mat{m} &= r \begin{pmatrix}
  1 & \tan\zeta\sin\phi_1\cos\phi_2 \\
  0 & 1
\end{pmatrix}
\begin{pmatrix} \sin\zeta & 0 \\
  0 & \cos\zeta
\end{pmatrix}
\begin{pmatrix}
  1 & 0 \\
  \tan\zeta\cos\phi_1 & 1
\end{pmatrix}
\\
&= r
\begin{pmatrix}
  {\tan\zeta\sin\zeta\sin\phi_1\cos\phi_1\cos\phi_2} &
  \sin\zeta \sin\phi_1\cos\phi_2 \\
  \sin\zeta \cos\phi_1 & \cos\zeta
\end{pmatrix} +
r
\begin{pmatrix}
  \sin\zeta & 0 \\
  0 & 0
\end{pmatrix}.
\end{aligned}\] } However, the usefulness of this approach does not lie
in a functional relation between matrix elements and mixing angles but
rather in the minimalistic picture it offers to generate zero matrix
elements or hierarchical elements and a complementary understanding of
both of them.

For small angles \(\rho \equiv \chi \sim \phi_1 \sim \phi_2 \ll 0\) we
can perform a Taylor expansion and find
\begin{equation} \label{eq:2fam-1zero}
\Mat{m} \sim \begin{pmatrix}
\rho^3 & \rho^2 \\
\rho - \frac{2}{3} \rho^3 & 1 - \frac{\rho^2}{2}
\end{pmatrix} + \mathcal{O}(\rho^4)
\quad \text{and} \quad
\Mat{m} \Mat{m}^\dag \sim \begin{pmatrix}
0 & \rho^2 \\ \rho^2 & 1
\end{pmatrix} + \mathcal{O}(\rho^4),
\end{equation}
which also justifies the discussion about hierarchical matrix elements
and a vanishing 1-1 entry in the Appendix of
Ref.~\cite{Hollik:2014jda}. Similarly, by setting \(\phi_2 \to 0\), we
insert one texture zero. Therefore, we see that there is a basis where,
\begin{equation}\label{eq:2flv2}
  \Mat{m} = r \begin{pmatrix}
    0 & \sin\chi \sin\phi_1 \\
    \sin\chi \cos\phi_1 & \cos\chi
  \end{pmatrix},
\end{equation}
and one reaches the same conclusion up to $\mathcal{O}(\rho^3)$ as under
the small angle approximation from
Eq. \eqref{eq:2fam-1zero}. Furthermore, we can put Eq.~\eqref{eq:2flv2}
into the form of a Cheng--Sher ansatz $|\Mat{m}_{ij}|\sim\sqrt{m_i m_j}$
\cite{Cheng:1987rs}, exploiting \(\sin(\chi) = \sqrt{1-\cos^2\chi} =
\sqrt{(1-\cos\chi) (1+\cos\chi)}\) (which works for \(\chi \in
[0,\tfrac{\pi}{2}]\)). Defining
\begin{equation}
m_1 = \tfrac{r}{\sqrt{2}} (1-\cos\chi) \qquad \text{and} \qquad
m_2 = \tfrac{r}{\sqrt{2}} (1+\cos\chi),
\end{equation}
we have together with \(\phi_1 = \tfrac{\pi}{4}\)
\begin{equation}
\Mat{m} = \begin{pmatrix}
  0 & \sqrt{m_1 m_2} \\ \sqrt{m_1 m_2} & \tfrac{1}{\sqrt{2}} (m_2 - m_1)
\end{pmatrix}.
\end{equation}

\section{Physical and unphysical zeros}
\label{sec:zeros}
It has become common use to introduce null mass matrix elements defined
as a certain ``ansatz'' and (or) put initially complex mass matrices
into hermitian form, arguing that weak basis transformations allow
them~\cite{Fritzsch:1977vd, Weinberg:1977hb, Branco:1988iq,
  Branco:1994jx, Branco:1999nb, Emmanuel-Costa:2016gdp}.  In this
section, we shall give a direct explanation of their origin in our
interpretation of mass matrices and comment on which texture zeros can
be called unphysical and which other can only be due to a physical
origin (e.g. a symmetries of the Lagrangian), reproducing the
conclusions already reached in the literature, see
Refs.~\cite{Branco:1988iq, Branco:1994jx, Branco:1999nb,
  Emmanuel-Costa:2016gdp}.

Consider the $n$ family case. As no right-handed charged currents have
been observed, right-handed transformations in family space are
unphysical; thus, giving a total of $n(n-1)$ arbitrary unphysical angles
per fermion sector. On the other hand, unitary transformations
preserving flavour invariance in the charged current interactions (weak
basis transformations) will contribute to this number with
$n(n-1)/2$. This set of angles, $3n(n-1)/2$ in total, is the one
responsible for producing unphysical zeros in a mass matrix or equal
mass matrix elements. The key difference from our approach with others
is that in a very simple manner one can track the consequences of making
a null element on the other matrix elements. By introducing these zeros,
the vector on the surface of the hypersphere gets aligned along certain
axes in flavour space as can be seen from the following subsection.

\subsection{Nearest-Neighbour-Interaction form}
For $n=3$, we have $9$ arbitrary and unphysical angles to which we can
assign any value. From Eq. \eqref{eq:sph-matrix-elems}, we see that
under the choice
\begin{equation}
        \phi_{2,4,6} = \frac{\pi}{2} \quad \text{and} \quad \phi_7 = 0 ,
\end{equation}
we easily generate the following well-known mass matrix, so called
Nearest-Neighbour-Interaction form \cite{Branco:1988iq}
\begin{equation}
        |\Mat{M}| = \begin{pmatrix}
                0 & A & 0 \\
                A' & 0 & B\\
                0 & B' & C
        \end{pmatrix},
\end{equation}
with
\begin{subequations}
\begin{align}
A &= R \sin\chi \sin\phi_1 \sin\phi_3 \sin\phi_5, \\
A' &= R \sin\chi \sin\phi_1 \sin\phi_3 \cos\phi_5, \\
B &= R \sin\chi \sin\phi_1 \cos\phi_3, \\
B' &= R \sin\chi \cos\phi_1, \\
C &= R \cos\chi.
\end{align}
\end{subequations}
We can then reexpress the spherical coordinates by the mass matrix
elements as
\begin{subequations}
\begin{align}
\tan\phi_5 &= \frac{A}{A'}, \\
\tan\phi_3 & = \sqrt{1 + \left(\frac{A}{A'}\right)^2}, \\
\tan\phi_1 & = \sqrt{1 + \left(1 + \left(\frac{A}{A'}\right)^2 \right)
  \left(\frac{A^2}{A'B}\right)^2} \frac{B}{B'}.
\end{align}
\end{subequations}
Moreover, we still have one more free angle by which we could choose $A'
= A$, that is $\phi_5 = \pi/4$. Although this would only hold for one of
the two mass matrices per fermion sector.

\subsection{Inclusion of complex phases}
The three unitary matrices giving rise to the weak basis transformations
imply a total of $[3n(n+1)-2]/2$ arbitrary (unphysical) complex phases.
For $n=3$ we have 17 free phases. In order to correctly introduce them
in the spherical mass matrix interpretation, we need to subtract the
number of phases gone when producing null mass matrix elements.  Take
for example our previous case, this implies having $17-8=9$ unphysical
phases left. The matrices have in total 10 complex phases. Through an
appropriate choice of phases, we are allowed to keep one independent
phase; which could also have been anticipated if after introducing the
textures zeros, one realises that only one linear combination of phases
remains in Eqs.\eqref{eq:xs-and-ys},
$\gamma=\delta_{21}+\delta_{33}-\delta_{31}-\delta_{23}$. Therefore, by
redefining them in such a way that only one $\delta_{21}$ survives we
get
\begin{equation} \label{eq:NNI}
  \Mat{M}_a = \begin{pmatrix}
    0 & A_a & 0 \\
    A'_a & 0 & B_a\\
    0 & B'_a & C_a
  \end{pmatrix}, \qquad \qquad
  \Mat{M}_b = \begin{pmatrix}
    0 & A_b e^{i\gamma} & 0 \\
    A_b e^{-i\gamma} & 0 & B_b\\
    0 & B'_b & C_b
  \end{pmatrix} ,
\end{equation}
giving a total of ten independent parameters in accordance with the ones
appearing in the mass basis. So we see that by relating the weak basis
transformations to the spherical mass matrix interpretation allows us to
directly write the matrix forms with all their redundancy now ripped
off.

\subsection{Hermiticity and texture zeros}
By demanding hermitian matrices, there is a cost one should pay, which
is on one hand 6 of the 9 angles have been employed while on the other,
12 of the 17 available phases have also been used. Therefore, the
introduction of further constraints as null matrix elements should be
limited to only 3 free angles and 5 complex phases. So equally
distributing three null mass matrix elements between two matrices is
impossible. That is, within the traditional approach, no parallel
structures with zero elements can be obtained via weak basis
transformations when hermiticity has been first
invoked.\footnote{Parallel structures are such matrix structures, where
  both matrices in the same fermion sector (quark or lepton) shares
  their matrix form.}  Within our approach this can also be
done. However, taking a look at Eqs.~\eqref{eq:sph-matrix-elems}, an
alternative scenario appears in which parallel structures \emph{seem}
to be allowed. In the following we will discuss the former scenario
(no-parallel structures) and then we will clarify the issue of the
alternative one (parallel structures).

Let us show it. For the former point, first apply the hermiticity
condition and thereafter the spherical mass matrix interpretation. The
space of the hypersphere now gets reduced from nine dimensions to only
six with the matrix elements given by
\begin{subequations}
  \begin{align}
\widetilde{m}_{11} &= R \sin\chi \sin\phi_1 \sin\phi_2 \sin\phi_3 \sin\phi_4, \\
\widetilde{m}_{12} &= R \sin\chi \sin\phi_1 \sin\phi_2 \sin\phi_3 \cos\phi_4,
\\
\widetilde{m}_{13} &= R \sin\chi \sin\phi_1 \sin\phi_2 \cos\phi_3, \\
\widetilde{m}_{22} &= R \sin\chi \sin\phi_1 \cos\phi_2, \\
\widetilde{m}_{23} &= R \sin\chi \cos\phi_1, \\
\widetilde{m}_{33} &= R \cos\chi.
\end{align}
\end{subequations}
Note, that for a hermitian matrix, one has an overcounting for the
Frobenius norm from the off-diagonal elements, so we define the mass
matrix as
\begin{equation}
\Mat{M} = \begin{pmatrix}
  \widetilde{m}_{11} &
  \tfrac{1}{\sqrt{2}} \widetilde{m}_{12} e^{i \delta_{12}} &
  \tfrac{1}{\sqrt{2}} \widetilde{m}_{13} e^{i \delta_{13}} \\
  \tfrac{1}{\sqrt{2}} \widetilde{m}_{12} e^{-i \delta_{12}} &
  \widetilde{m}_{22} &
  \tfrac{1}{\sqrt{2}} \widetilde{m}_{23} e^{i \delta_{23}} \\
  \tfrac{1}{\sqrt{2}} \widetilde{m}_{13} e^{-i \delta_{13}} &
  \tfrac{1}{\sqrt{2}} \widetilde{m}_{23} e^{-i \delta_{23}} &
  \widetilde{m}_{33}
\end{pmatrix}.
\end{equation}

In this sense, we have now five angles from which two may correspond to
the singular values and the other three allow to introduce texture
zeros. Nevertheless, in total we have no more than three free angles for
both matrices. So following this, we can produce the next kind of
no-parallel weak basis matrices,
\begin{equation}
  \Mat{M}_a = \begin{pmatrix}
    0 & A_a & 0 \\
    A_a & B_a & C_a\\
    0 & C_a & D_a
  \end{pmatrix}, \qquad \qquad
  \Mat{M}_b = \begin{pmatrix}
    A_b & B_b e^{i\beta} & 0 \\
    B_b e^{-i\beta} & C_b & D_b\\
    0 & D_b & E_b
  \end{pmatrix},
\end{equation}
thus reaching the traditional
conclusions~\cite{Branco:1999nb}. Apparently, we have chosen $\phi_4^{a}
= 0$, $\phi_3^{a(b)} = \tfrac{\pi}{2}$. First of all, there is no
\emph{physical meaning} attached to any of those zeros in a certain weak
basis like the one we have singled out here. We have to reduce the
number of free parameters to ten---how this is achieved should have no
influence on the observable physics. Second, there can be no parallel
structures for hermitian matrices with only one complex phase. However,
with \(\phi_4^{b} = 0\), one either has to introduce an additional phase
or one can construct a \emph{prediction} of one of the SM parameters in
terms of the others. This is only valid by \emph{adhoc} assumptions or
proposing a kind of flavour symmetry. In the latter case, there is, of
course, a physical meaning associated with it; see for
example~\cite{deMedeirosVarzielas:2017sdv}.

One remark about alternative scenarios and possible loopholes in our
interpretation: Notice that if we had considered
$\phi^b_2={\tfrac{\pi}{2}}$ in the second matrix we could have found a
parallel structure. And moreover, for $\phi_3^b =0$ and $\phi^a_3 = 0$
plus an adequate initial reordering of the matrix entries, we could have
found another parallel structure. Therefore, it \emph{seems} that we can
indeed build parallel structures with more than three independent
texture zeros together with hermiticity. What seems to be wrong? Both
alternative scenarios reach a weak basis with less than ten arbitrary
parameters. But this contradicts our interpretation on the angles which
corresponded to the freedom in the weak basis transformations (one
cannot have a \emph{weak basis} with less than ten arbitrary
parameters). Hence, the alternative scenarios are not valid within the
approach.

\subsection{Deviations from hermiticity in the
  Nearest-Neighbour-Interaction form}

From the point of view of our approach and the traditional ones,
producing the Nearest-Neighbour-Interaction form together with an
hermitian matrix, is impossible. However, from Eq.\eqref{eq:NNI}, we
could work out the deviations from hermiticity if we work in the small
angle approximation (further results about small angles in the next
section). With the assignment
\[
\phi^{a(b)}_7 = 0,
\phi_{2,4,6}^{a(b)} = \frac{\pi}{2}, \phi_{5}^{a} = \frac{\pi}{4},
\phi_{1}^{a} = \frac{\pi}{4} + \varepsilon_1^a, \text{ and }
\phi_{1,5}^{b} = \frac{\pi}{4} + \varepsilon_{1,5}^b,
\]
where $\varepsilon_{j} \ll 1$, we get the following:
\begin{small}
\begin{align}
  \Mat{M}_a \simeq \begin{pmatrix}
    0 & A_a & 0 \\
    A_a & 0 & B_a\\
    0 & B_a & C_a
  \end{pmatrix} + \varepsilon_1^a \begin{pmatrix}
    0 & 0 & 0 \\
    0 & 0 & B_a\\
    0 & -B_a & 0
  \end{pmatrix} +\mathcal{O}\left( (\varepsilon_5^a)^2 \right) ,\\
  \Mat{M}_b \simeq \begin{pmatrix}
    0 & A_b e^{i\beta} & 0 \\
    A_b e^{-i\beta} & 0 & B_b\\
    0 & B_b & C_b
  \end{pmatrix}  + \varepsilon_1^b \begin{pmatrix}
    0 & 0 & 0 \\
    0 & 0 & B_b\\
    0 & -B_b & 0
  \end{pmatrix} + \varepsilon_5^b \begin{pmatrix}
    0 & A_b e^{i\beta} & 0 \\
    -A_b e^{-i\beta}& 0 & 0\\
    0 & 0 & 0
  \end{pmatrix} +\mathcal{O}\left( (\varepsilon_{1,5}^b)^2 \right).
\end{align}
\end{small}
It can be readily seen how the presence of the small deviations helps to
the counting of ten free parameters within the weak basis. This approach
reduces from four to three parameters, as previously
used~\cite{Branco:1992ba, Branco:2010tx}, to measure the deviations from
hermiticity. It is a straightforward task to determine that this set of
parameters reproduce both the masses and the mixing in the quark sector.

\section{Majorana neutrinos}\label{sec:majorana}

Massive neutrinos are not part of the renormalisable Standard
Model. There is, however, one single operator at dimension five that can
generate very small neutrino masses for the left-handed neutrinos only
\cite{Weinberg:1979sa}, without introducing right-handed neutrinos. The
UV-completion of this operator will reveal some new physics at the scale
\(\Lambda_\text{NP}\). This operator requests the resulting mass matrix
to be of the Majorana type, meaning complex but symmetric. It is a
gauge- and Lorentz-invariant construction:
\begin{equation}\label{eq:dim5op}
  \Lag_5 = \frac{1}{2} \frac{c_{\alpha\beta}}{\Lambda_\text{NP}}
  \left({\bar L}^c_{L\alpha}\widetilde{H}^* \right)
  \left(\widetilde{H}^\dagger {L_{L\beta}} \right) + \hc,
\end{equation}
where $L_L = (\nu_L, e_L)^T$ and $H=(H^+,H^0)^T$ are the left-handed
lepton and the Higgs doublet of the SM, respectively; we follow the
usual notation for the charged conjugated Higgs field as
$\widetilde{H} = i \sigma_2 H^*$. The coefficients $c_{\alpha\beta}$
are arbitrary numbers, but supposed to be \(\mathcal{O}(1)\) numbers
or show some rather mild hierarchy which is imprinted in the neutrino
mass spectrum. The whole operator is suppressed by the new physics
scale, $\sim 1/\Lambda_{NP}$, which can be \(\mathcal{O}(10^{10\ldots
  14}\,\GeV)\).

We want to study the different zero elements that could arise from weak
basis transformations. The flavour group for the lepton sector is
\begin{equation}
\mathcal{G}_F \supset U_L^\ell(3) \times U_R^e(3).
\end{equation}
The above group of transformations can be used to diagonalise the
charged lepton mass matrix. In this weak basis, which we could call the
charged lepton basis, the symmetrical mass matrix of neutrinos gets
diagonalised by the Pontecorvo--Maki--Nakagawa--Sakata (PMNS) matrix. So
we immediately reach the conclusion that as no freedom is left to still
make weak basis transformations, any texture zero in the neutrino mass
matrix will be physical as long as we are in the charged lepton basis.

\subsection{Weak bases}
Let us consider those weak bases where the charged lepton mass matrices
are still non diagonal. This discussion not only reproduces some known
facts, as of Ref.~\cite{Branco:2007nn}, but, if extended, may provide
further observations. From the six free angles, we can choose four of
them as $\phi^{e}_7 = 0$ and $\phi_{2,4,6}^{e} = \frac{\pi}{2}$ in the
spherical mass matrix interpretation, to get \eg
\begin{equation}
\Mat{M}_e = \begin{pmatrix}
  0 & A_a e^{-i\delta} & 0 \\
  A'_e e^{i\delta} & 0 & B_e\\
  0 & B'_e & C_e
\end{pmatrix}.
\end{equation}
The neutrino mass matrix, however, has to be symmetric. We change the
notation slightly and perform a renaming \(\phi \to \omega\) in the
angles to show the difference. Hence, we have the following entries
\begin{subequations}
  \begin{align}
    \widetilde{m}_{11}^\nu &=
    R^\nu \sin\chi^\nu \sin\omega_1^\nu \sin\omega_2^\nu \sin\omega_3^\nu \sin\omega_4^\nu, \\
    \widetilde{m}_{12}^\nu &=
    R^\nu \sin\chi^\nu \sin\omega_1^\nu \sin\omega_2^\nu \sin\omega_3^\nu \cos\omega_4^\nu, \\
    \widetilde{m}_{13}^\nu &=
    R^\nu \sin\chi^\nu \sin\omega_1^\nu \sin\omega_2^\nu \cos\omega_3^\nu, \\
    \widetilde{m}_{22}^\nu &= R^\nu \sin\chi^\nu \sin\omega_1^\nu \cos\omega_2^\nu, \\
    \widetilde{m}_{23}^\nu &= R^\nu \sin\chi^\nu \cos\omega_1^\nu, \\
    \widetilde{m}_{33}^\nu &= R^\nu \cos\chi^\nu,
\end{align}
\end{subequations}
of the complex symmetric matrix
\begin{equation}
\Mat{M}^\nu = \begin{pmatrix}
  \widetilde{m}^\nu_{11} e^{i \varphi^\nu_{11}} &
  \tfrac{1}{\sqrt{2}} \widetilde{m}^\nu_{12} e^{i \varphi^\nu_{12}} &
  \tfrac{1}{\sqrt{2}} \widetilde{m}^\nu_{13} e^{i \varphi^\nu_{13}} \\
  \tfrac{1}{\sqrt{2}} \widetilde{m}^\nu_{12} e^{i \varphi^\nu_{12}} &
  \widetilde{m}^\nu_{22} e^{i \varphi^\nu_{22}} &
  \tfrac{1}{\sqrt{2}} \widetilde{m}^\nu_{23} e^{i \varphi^\nu_{23}} \\
  \tfrac{1}{\sqrt{2}} \widetilde{m}^\nu_{13} e^{i \varphi^\nu_{13}} &
  \tfrac{1}{\sqrt{2}} \widetilde{m}^\nu_{23} e^{i \varphi^\nu_{23}} &
  \widetilde{m}^\nu_{33} e^{i \varphi_{33}}
\end{pmatrix}.
\end{equation}

From the two remaining unphysical degrees of freedom, we can induce
several texture zeros in the neutrino masses, \eg with $\omega_4^\nu =
0$ and $\omega_3^\nu = \tfrac{\pi}{2}$ we find
\begin{equation}
  \Mat{M}_\nu = \begin{pmatrix}
    0 & A_\nu e^{-i\alpha_1}& 0 \\
    A_\nu e^{-i\alpha_1} & B_\nu & C_\nu e^{-i\alpha_2}\\
    0 & C_\nu e^{-i\alpha_2}& D_\nu
  \end{pmatrix} .
\end{equation}
It is outside the scope of this work to provide an exhaustive list of
different weak basis matrix forms. Therefore, the take-home message lies
in the simplicity of the spherical mass matrix interpretation on
studying matrices in different weak bases.

\subsection{Phenomenological application: the Altarelli--Feruglio model}

The charged lepton basis is ideal to get further insights into the
masses or mixing of neutrinos, as everything is extracted from their
mass matrix. In this regard, the famous Altarelli--Feruglio model
provides us with a good example
\cite{Altarelli:2005yp,Altarelli:2005yx}.  The model implements the
$A_4$ non-Abelian and discrete symmetry group inside a Frogatt--Nielsen
framework. It naturally implies tribimaximal mixing (TBM) for the PMNS
matrix \cite{Harrison:2002er}:
\begin{equation}
  \Mat{U}_\text{TBM} = \begin{pmatrix}
    \sqrt{\frac{2}{3}} & \sqrt{\frac{1}{3}} & 0 \\
    -\sqrt{\frac{1}{6}} & \sqrt{\frac{1}{3}} & -\sqrt{\frac{1}{2}}  \\
    -\sqrt{\frac{1}{6}} & \sqrt{\frac{1}{3}} & \sqrt{\frac{1}{2}}
  \end{pmatrix}.
\end{equation}
Its weak point, however, is that the reactor mixing angle is predicted
to be exactly zero, $\theta_{13}^\nu = 0$, so it is in the meantime
excluded by experimental observation \cite{Ahn:2012nd, An:2012eh,
  Abe:2013hdq}.  Nevertheless, the main ingredient of the model, the
underlying tribimaximal mixing, still can be relevant for a partial
diagonalisation. The fact, that neutrino masses are less hierarchical
than charged fermion masses, suggest a more democratic flavour pattern,
which is related to tribimaximal mixing.

Three main features characterize the Altarelli--Feruglio mass matrix:
$m^\nu_{12} = m^\nu_{13}$, $m^\nu_{22} = m^\nu_{33}$, and $m^\nu_{22} =
-2 m^\nu_{12}$. Under the spherical mass matrix interpretation we look
for the consequences of implementing them starting from the charged
lepton basis.

We assume the following assignment of the real matrix elements:
\begin{subequations}
  \begin{align}
    \widetilde{m}_{11}^\nu &=
    R^\nu \sin\chi^\nu \sin\omega_1^\nu \sin\omega_2^\nu \cos\omega_3^\nu, \\
    \widetilde{m}_{12}^\nu &=
    R^\nu \sin\chi^\nu \sin\omega_1^\nu \sin\omega_2^\nu \sin\omega_3^\nu \sin\omega_4^\nu, \\
    \widetilde{m}_{13}^\nu &=
    R^\nu \sin\chi^\nu \sin\omega_1^\nu \sin\omega_2^\nu \sin\omega_3^\nu \cos\omega_4^\nu, \\
    \widetilde{m}_{22}^\nu &= R^\nu \sin\chi^\nu \sin\omega_1^\nu \cos\omega_2^\nu, \\
    \widetilde{m}_{23}^\nu &= R^\nu \sin\chi^\nu \cos\omega_1^\nu, \\
    \widetilde{m}_{33}^\nu &= R^\nu \cos\chi^\nu.
\end{align}
\end{subequations}

The equality of $\widetilde{m}_{12}^\nu = \widetilde{m}_{13}^\nu$
implies a basis choice in which $\omega_4^\nu = \tfrac{\pi}{4}$. On the
other hand, with $\widetilde{m}_{33}^\nu = \widetilde{m}_{22}^\nu$ one
needs $\tan \chi^\nu \geq 1$ and thus $\chi^\nu \in
[\tfrac{\pi}{4},\tfrac{\pi}{2})$. Note how one may identify the
particular choice of the elements with a particular orientation of the
mass vector in the flavour basis. Last, we require
$\widetilde{m}_{22}^\nu = -2 \widetilde{m}_{12}^\nu$ and see that it is
only fulfilled with $\omega_3^\nu = \tfrac{3\pi}{2}$ and \(\omega_2^\nu
= \tfrac{5 \pi}{4}\). Under these conditions one gets the following mass
matrix,
\begin{equation}
        |\Mat{M}_\nu| = \begin{pmatrix}
    0 & a^\nu & a^\nu \\
    a^\nu & -2a^\nu & b^\nu\\
    a^\nu & b^\nu & -2a^\nu
  \end{pmatrix} ,
\end{equation}
where we have $a^\nu =\frac{R^\nu}{2\sqrt{2}}\sin\chi^\nu
\sin\omega_1^\nu$ and $b^\nu = \frac{R^\nu}{\sqrt{2}} \sin\chi^\nu
\cos\omega_1^\nu$, and the relation
\begin{equation}
\tan\chi^\nu \sin\omega_1^\nu = - \sqrt{2}.
\end{equation}
Notice that it does not reproduce the full Altarelli--Feruglio mass
matrix (e.g. $m_{11}^\nu \neq 0$). Therefore, we expect a deviation from
tribimaximal mixing, which is actually required by experiment. The
vanishing 1-1 element in our case is a direct consequence of the
spherical mass matrix interpretation as the individual elements are not
fully independent.

Let us decompose the mass matrix into a democratic part and a remainder
which only has 2-3 mixing
\begin{equation} \label{eq:numassdecomp}
        |\Mat{M}_\nu| = \begin{pmatrix}
    a^\nu & a^\nu & a^\nu \\
    a^\nu & a^\nu & a^\nu\\
    a^\nu & a^\nu & a^\nu
  \end{pmatrix} +
  \begin{pmatrix}
    -a^\nu & 0 & 0 \\
    0 & -3a^\nu & b^\nu-a^\nu\\
    0 & b^\nu-a^\nu & -3a^\nu
  \end{pmatrix}.
\end{equation}
The first term gets diagonalised by the tribimaximal mixing
matrix. After that, we have
\begin{equation} \label{eq:numass'}
  |\Mat{M}'_\nu| = \begin{pmatrix}
    \frac{1}{3} (-6a^\nu+b^\nu) & \frac{\sqrt{2}}{3} (3a^\nu-b^\nu) & 0 \\
    \frac{\sqrt{2}}{3} (3a^\nu-b^\nu) & \frac{2b^\nu}{3} & 0\\
    0 & 0 & -2a^\nu-b^\nu
  \end{pmatrix} ,
\end{equation}
which still requires a further diagonalisation. This, however, can be
done trivially. The full PMNS matrix is then given by the initial
tribimaximal mixing matrix, corrected with the diagonalisation of
Eq.~\eqref{eq:numass'}. Since there are furthermore only two independent
parameters, \(a^\nu\) and \(b^\nu\), the mass spectrum as well as the
neutrino mixing matrix can be fully determined by a fit to the
experimentally known mass squared differences only. With the most recent
results of \cite{Esteban:2016qun}\footnote{Similar results can be found
  in other sources like~\cite{deSalas:2017kay}. We only perform a proof
  of principle here and also do no error analysis, just to see whether
  we roughly get the right numbers.}
\[
\Deltaup m^2_{21} = 7.40 \times 10^{-5} \, \eV^2,
\qquad \text{and} \qquad
\Deltaup m^2_{31} = 2.494 \times 10^{-3} \, \eV^2,
\]
we obtain, assuming normal hierarchy and ignoring the errorbars, we get
two real and positive solutions for \(a^\nu\) and \(b^\nu\), that are
very close
\begin{subequations}
\begin{align}
a^\nu &\in \{ 0.0127, 0.0138\}\,\eV, \qquad \text{and} \\
b^\nu &\in \{ 0.0274, 0.0257\}\,\eV.
\end{align}
\end{subequations}
This determines the neutrino mass spectrum to be for the two solutions
\begin{equation}
m_3^\nu = \{ 0.0527, 0.0533 \} \,\eV, \quad
m_2^\nu = \{ 0.0190, 0.0205 \} \, \eV, \quad
m_1^\nu = \{ 0.0169, 0.0186 \} \, \eV,
\end{equation}
and the PMNS matrix for both the cases
\begin{equation}
\left|\Mat{U}_\text{PMNS}\right| = \left\{ \begin{pmatrix}
  0.727 & 0.686 & 0 \\
  0.485 & 0.514 & 0.707 \\
  0.485 & 0.514 & 0.707
\end{pmatrix},
 \begin{pmatrix}
  0.724 & 0.690 & 0 \\
  0.488 & 0.512 & 0.707 \\
  0.488 & 0.512 & 0.707
\end{pmatrix} \right\}.
\end{equation}
Apparently, this PMNS matrix cannot describe the true neutrino
phenomenology, which is also not surprising: the Altarelli--Feruglio
models were invented to predict a \(\theta^\nu_{13} = 0\), and staying
within the underlying pattern for the mass matrix, we cannot generate a
non-vanishing entry there.

It is, however, astonishingly simple to correct for a non-vanishing 1-3
mixing. The Altarelli--Feruglio matrix cannot have a 1-3 mixing: from
Eq.~\eqref{eq:numassdecomp}, we see that the non-democratic part of the
mass matrix does not mix the first and third generation. We can
nevertheless accomodate for it by a small \emph{misalignment} of the two
elements \(\widetilde{m}^\nu_{12}\) and \(\widetilde{m}^\nu_{13}\),
simply with the choice \(\omega^\nu_4 = \tfrac{\pi}{4} + \varepsilon\),
leading to a corrected mass matrix
\begin{equation}
        |\Mat{M}_\nu| = \begin{pmatrix}
    0 & a^\nu + \delta^\nu & a^\nu - \delta^\nu \\
    a^\nu + \delta^\nu & -2a^\nu & b^\nu\\
    a^\nu -\delta^\nu & b^\nu & -2a^\nu
  \end{pmatrix} + \mathcal{O}(\varepsilon^2),
\end{equation}
with \(\delta^\nu = a^\nu \varepsilon\). With \(\delta^\nu\), we have a
handle on \(\theta^\nu_{13}\) and in order to generate
\(\sin\theta^\nu_{13} \approx 0.15\), we find \(\delta^\nu = 0.005\,\eV\) and
one set of solutions with
\begin{equation}
a^\nu = 0.0126\,\eV, \qquad\text{and}\qquad b^\nu = 0.0263\,\eV,
\end{equation}
resulting in a slightly modified mass spectrum
\begin{equation}
m^\nu_3 = 0.0526 \,\eV, \quad m^\nu_2 = 0.0187 \,\eV \quad \text{and}
\quad m^\nu_1 = 0.0166 \,\eV.
\end{equation}
This na\"ive correction still has some tension in comparison with the
global fit values of the PMNS matrix. We find
\begin{equation}
\left|\Mat{U}_\text{PMNS}\right| = \begin{pmatrix}
  0.696 & 0.702 & 0.150 \\
  0.398 & 0.551 & 0.733 \\
  0.598 & 0.451 & 0.663
\end{pmatrix}.
\end{equation}

We nowadays have strong hints of \CP{} violation in neutrino
oscillations, besides the fact that the third mixing angle is definitely
non-zero. Furthermore, recent global fits tend towards a rather maximal \CP-phase in
the Standard Parametrisation (\(\delta_{CP} =
{234^{+43}_{-31}}\,^\circ\)~\cite{Esteban:2016qun}), which is compatible
with \(\delta_{CP} \approx -90\,^\circ\). TBM mixing is thus ruled out and
the Altarelli--Feruglio model has to be adjusted for this, including
\CP{} violation. This easily can be accommodated within the approach
presented above. Let us consider an \emph{imaginary} perturbation,
\(\omega^\nu_4 = \tfrac{\pi}{4} + \im \varepsilon\), and thus
\(\sin(\omega^\nu_4) \approx (1 + \im\epsilon)\sqrt{2}\), we can simply multiply
\(\delta^\nu\) with a maximal complex phase \(e^{\im \pi/2}\). Keeping
\(\delta^\nu = 0.005\), to achieve a large \(\sin\theta_{13}^\nu \approx
0.15\), this modifies slightly the mass eigenvalues. Hence, to reproduce
the proper \(\Deltaup m^2\), we have to refit the \(a^\nu\) and
\(b^\nu\) parameters and find
\begin{equation}
a^\nu = 0.0127\,\eV, \qquad\text{and}\qquad b^\nu = 0.0285\,\eV,
\end{equation}
and correspondingly
\begin{equation}
m^\nu_3 = 0.0528 \,\eV, \quad m^\nu_2 = 0.0193 \,\eV \quad \text{and}
\quad m^\nu_1 = 0.0172 \,\eV.
\end{equation}
The PMNS matrix now has a complex phase and is given by
\begin{equation}
\Mat{U}_\text{PMNS} = \begin{pmatrix}
  0.742 & 0.668 & -0.00715 + 0.148 \im \\
  -0.463 + 0.101 \im & 0.524 + 0.0456 \im & -0.696 - 0.0673 \\
  -0.463 - 0.101 \im & 0.524 - 0.0456 \im & 0.699
\end{pmatrix}.
\end{equation}
This has surprisingly a \CP-phase \(\delta_{CP} = -0.485 \pi\) in
accordance with the global fit.

\section{Small angles and hierarchies} \label{sec:smallangles}

Generically, it is believed that any kind of hierarchy in the
eigenvalues (singular values) of a mass matrix has to be already coded
in the hierarchical structure of the individual matrix elements, as was
proposed by Froggatt and Nielsen \cite{Froggatt:1978nt}
\begin{equation}\label{eq:FN}
  - \mathcal{L}_\text{FN} = \sum_{n,\psi} \bar{\psi}_{L,i} \psi_{R,j} H \lambda_{ij}^\psi \left( \frac{\varphi}{\Lambda}\right)^{n_{ij}} + \hc,
\end{equation}
where \(\psi_i\) are generic fermions with \(i=1,2,3\) counting the
number of generations, \(H\) being the SM Higgs doublet breaking
electroweak gauge symmetry and \(\varphi\) a flavon field breaking the
continous and global flavour symmetry. The flavour symmetry is assumed
to be an Abelian \(\U(1)_F\) global symmetry and the ``coupling
constants'' \(\lambda_{ij}^\psi\) are supposed to be \emph{arbitrary}
\(\mathcal{O}(1)\) numbers, where the additional scale \(\Lambda\)
refers to a larger scale at which new degrees of freedom are integrated
out. So the final ``Yukawa couplings'' as effective couplings of the SM
fermions to the SM Higgs are given by
\begin{equation}
Y_{ij}^f = \lambda_{ij}^f \left(
  \frac{\langle \varphi \rangle}{\Lambda} \right)^{n_{ij}},
\end{equation}
with \(n_{ij} \in \mathds{N}\) being the sum of the corresponding
\(\U(1)_F\) charges. The hierarchical fermion masses are then encoded in
powers of a small parameter \(\varepsilon = \langle \varphi \rangle /
\Lambda\). As numerical example: take \(\Lambda\) to be \(10\,\TeV\) and
\(\langle \varphi\rangle\) to be of the electroweak scale \(\sim
100\,\GeV\), then \(\varepsilon \simeq 10^{-2}\).

Therefore, apparently, a hierarchical matrix configuration can only be
attached to the idea of a complicated mechanism fully responsible for
it. The art of finding a viable UV-completion of this model typically
leads to vastly extended sets of matter and scalar fields and may not be
called aesthetic. In the following, we explore a different route to
arrive at a very similar suppression of small numbers by high powers
employing the spherical mass matrix interpretation. The small numbers
then arise from a small misalignment of the mass vector with respect to
the underlying flavour basis.

Let us consider all the angles very close to zero, so the actual vector
in the \(9\)-dimensional space points along the
\(m_{33}\)-axis. Surprisingly, one finds immediately
Froggatt--Nielsen-like structures. Let us take all angles to be of the
same order, say \(\varepsilon \equiv \chi \sim \phi_k^{a(b)} \ll 1\),
and we get
\begin{equation}\label{eq:example}
  |\Mat{M}| \sim R
  \begin{pmatrix}
    \varepsilon^8 & \varepsilon^7 & \varepsilon^6 \\
    \varepsilon^5 & \varepsilon^4 & \varepsilon^3 \\
    \varepsilon^2 & \varepsilon & 1
  \end{pmatrix},
\end{equation}
\emph{without} referring to a Froggatt--Nielsen (FN) mechanism of
Eq.~\eqref{eq:FN}. Notice also, that the pattern of
Eq.~\eqref{eq:example} is not unique and, moreover, there is no reason
not to treat individual angles individually. A very obvious
transformation of this kind is \(\chi \to \chi - \frac{\pi}{2}\), then
the 3-3 element becomes \(\sim\varepsilon\) and the power of epsilons is
reduced by one on the other elements.  The key part in this construction
is, that---depending on the alignment in the abstract high dimensional
space---hierarchies can be generated by the choice of the basis and a
hierarchical basis as suggested by the FN mechanism does not imply
hierarchy of new physics scales.  Finally, the relevant object to
construct the mixing matrix is the left-hermitian product
\begin{equation}
| \Mat{M} \Mat{M}^\dag | \sim R^2 \begin{pmatrix}
\varepsilon^{12} + \mathcal{O}(\varepsilon^{14}) &
\varepsilon^9 + \mathcal{O}(\varepsilon^{11}) &
\varepsilon^6 + \mathcal{O}(\varepsilon^8) \\
\varepsilon^9 + \mathcal{O}(\varepsilon^{11}) &
\varepsilon^6 + \mathcal{O}(\varepsilon^8) &
\varepsilon^3 + \mathcal{O}(\varepsilon^5) \\
\varepsilon^6 + \mathcal{O}(\varepsilon^8) &
\varepsilon^3 + \mathcal{O}(\varepsilon^5) &
1 + \varepsilon^2 + \mathcal{O}(\varepsilon^4)
\end{pmatrix}
\end{equation}
which shows a \emph{strong} hierarchical structure.

Now, let us give a twist to the story. As previously noted, hierarchical
mass ratios are a direct consequence of \emph{only} two small angles, if
we assign spherical coordinates to the singular values in a similar
manner. Accordingly, there is no need to have all the eight angles as
small numbers, $\phi_i,\chi \ll 1$. So to produce mass hierarchies, we
actually do not need such a very strong suppression in all matrix
elements.  It is sufficient to have the following kind of mild
hierarchical structures:
\begin{equation} \label{eq:example2}
|\Mat{M}| \sim R
\begin{pmatrix}
\varepsilon^2 & \varepsilon^2 & \varepsilon^2 \\
\varepsilon^2 & \varepsilon^2 & \varepsilon \\
\varepsilon & \varepsilon & 1
\end{pmatrix} \qquad \Rightarrow \qquad
| \Mat{M} \Mat{M}^\dag | \sim R^2 \begin{pmatrix}
\varepsilon^{4} & \varepsilon^3 & \varepsilon^2 \\
\varepsilon^3  & \varepsilon^2  & \varepsilon  \\
\varepsilon^2  & \varepsilon  & 1 + \varepsilon^2
\end{pmatrix}.
\end{equation}

\section{Conclusions}\label{sec:concl}

We have introduced a new and innovative interpretation of the fermion
mass matrix elements in the SM. This interpretation allows
cross-relations to weak basis transformations. The key element is found
in one of the matrix invariants involving the trace of the
left-hermitian product. Its equation simultaneously describes the
surface of a nine dimensional hypersphere with its radius equal to the
Frobenius norm of the mass matrix. This interpretation is trivially not
constrained to three families but applies to all \(n \times n\) mass
matrices. The idea of assigning to each matrix element a basis of
spherical coordinates, provides a framework to correlate their
magnitudes in a very simple manner. Moreover, it can be seen from this
approach that individual matrix elements cannot be set to zero without
affecting also others. There are eight angles needed in the spherical
mass matrix interpretation which can be furthermore related to the weak
basis angles and the singular values of the mass matrices. Therefore,
this interpretation also allows to relate introduction of null elements,
so called texture zeros, to a geometrical alignment in the underlying
flavour space.

A very compelling application of this approach has been found in the
neutrino sector. The main characteristics of the neutrino mass matrix in
the Altarelli--Feruglio can be mapped to a set of conditions for the
angles in the spherical mass matrix interpretation. Within the
Altarelli--Feruglio model, we have been able to fully determine the mass
spectrum as well as the neutrino mixing matrix. By virtue of a small
correction in terms of a perturbation of one of the angles, we
furthermore could reproduce a large reactor angle which is initially
zero in that model. Moreover, with a purely imaginary perturbation, the
value of the Dirac \CP-phase in the PMNS matrix turns out to be close to
the value favoured by the global fit.

In the same approach, with a small misalignment, it is easy to reproduce
Froggatt--Nielsen like patterns for hierarchical mass matrices without
the need of introducing a new physics scale or a complicated
UV-completion for such suppression. Nevertheless, the mechanism behind
this misalignment stays unclear at this stage. The spherical mass matrix
interpretation is not to be seen as a dynamical model of flavour but
shall rather help to simplify model assumptions behind such models. With
the interpretation of aligning or misaligning individual mass matrix
elements with a certain direction in flavour space, it might be possible
to draw conclusions going further than texture zeros. We want to remind,
that actually the flavour bases for the up and down sector are not fully
independent in the spherical mass matrix interpretation and thus, in a
more deeper analysis, relations between up- and down-type fermion masses
shall be revealed.

\section*{Acknowledgements}
We acknowledge useful discussions with L.~Diaz-Cruz and thank C.~Bonilla
and M.~Spinrath for a careful reading of the manuscript and their
comments on it. Moreover, we want to thank the anonymous referee for the
suggestions improving the paper. WGH acknowledges support from the DESY
Fellowship Program.  UJSS acknowledges support from a DAAD One-Year
Research Grant.  Furthermore, UJSS wants to thank the DESY Theory Group
for hospitality, when this project was initiated, and WGH wants to thank
the TTP at KIT for hospitality, when it was finalised.

\bibliographystyle{utcaps}
\bibliography{hierarchies}

\providecommand{\href}[2]{#2}\begingroup\raggedright\begin{thebibliography}{10}\small

\bibitem{Hollik:2014jda}
W.~G. Hollik and U.~J. Saldana-Salazar, ``{The double mass hierarchy pattern:
  simultaneously understanding quark and lepton mixing}'',
  \href{http://dx.doi.org/10.1016/j.nuclphysb.2015.01.019}{{\em Nucl. Phys.}
  {\bfseries B892} (2015) 364--389},
\href{http://arxiv.org/abs/1411.3549}{{\ttfamily arXiv:1411.3549 [hep-ph]}}.

\bibitem{Saldana-Salazar:2015raa}
U.~J. Saldana-Salazar, ``{The flavor-blind principle: A symmetrical approach to
  the Gatto-Sartori-Tonin relation}'',
  \href{http://dx.doi.org/10.1103/PhysRevD.93.013002}{{\em Phys. Rev.}
  {\bfseries D93} no.~1, (2016) 013002},
\href{http://arxiv.org/abs/1509.08877}{{\ttfamily arXiv:1509.08877 [hep-ph]}}.

\bibitem{Diaz-Cruz:2016pmm}
J.~L. Diaz-Cruz, W.~G. Hollik, and U.~J. Saldana-Salazar, ``{A bottom-up
  approach to the strong CP problem}'',
\href{http://arxiv.org/abs/1605.03860}{{\ttfamily arXiv:1605.03860 [hep-ph]}}.

\bibitem{Saldana-Salazar:2016pms}
U.~J. Saldana-Salazar, ``{Fermion masses as mixing parameters in the SM}'',
  \href{http://dx.doi.org/10.1088/1742-6596/761/1/012045}{{\em J. Phys. Conf.
  Ser.} {\bfseries 761} no.~1, (2016) 012045},
\href{http://arxiv.org/abs/1608.05341}{{\ttfamily arXiv:1608.05341 [hep-ph]}}.

\bibitem{Saldana-Salazar:2016hxb}
U.~J. Saldana-Salazar, ``{A principle for the Yukawa couplings}'',
  \href{http://dx.doi.org/10.1088/1742-6596/761/1/012064}{{\em J. Phys. Conf.
  Ser.} {\bfseries 761} no.~1, (2016) 012064},
\href{http://arxiv.org/abs/1607.07898}{{\ttfamily arXiv:1607.07898 [hep-ph]}}.

\bibitem{Fritzsch:1977vd}
H.~Fritzsch, ``{Weak Interaction Mixing in the Six - Quark Theory}'',
\href{http://dx.doi.org/10.1016/0370-2693(78)90524-5}{{\em Phys. Lett.}
  {\bfseries 73B} (1978) 317--322}.

\bibitem{Weinberg:1977hb}
S.~Weinberg, ``{The Problem of Mass}'',
\href{http://dx.doi.org/10.1111/j.2164-0947.1977.tb02958.x}{{\em Trans. New
  York Acad. Sci.} {\bfseries 38} (1977) 185--201}.

\bibitem{Branco:1988iq}
G.~C. Branco, L.~Lavoura, and F.~Mota, ``{Nearest Neighbor Interactions and the
  Physical Content of Fritzsch Mass Matrices}'',
\href{http://dx.doi.org/10.1103/PhysRevD.39.3443}{{\em Phys. Rev.} {\bfseries
  D39} (1989) 3443}.

\bibitem{Ramond:1993kv}
P.~Ramond, R.~G. Roberts, and G.~G. Ross, ``{Stitching the Yukawa quilt}'',
  \href{http://dx.doi.org/10.1016/0550-3213(93)90159-M}{{\em Nucl. Phys.}
  {\bfseries B406} (1993) 19--42},
\href{http://arxiv.org/abs/hep-ph/9303320}{{\ttfamily arXiv:hep-ph/9303320
  [hep-ph]}}.

\bibitem{Branco:1994jx}
G.~C. Branco and J.~I. Silva-Marcos, ``{NonHermitian Yukawa couplings?}'',
\href{http://dx.doi.org/10.1016/0370-2693(94)91069-3}{{\em Phys. Lett.}
  {\bfseries B331} (1994) 390--394}.

\bibitem{Branco:1999nb}
G.~C. Branco, D.~Emmanuel-Costa, and R.~Gonzalez~Felipe, ``{Texture zeros and
  weak basis transformations}'',
  \href{http://dx.doi.org/10.1016/S0370-2693(00)00193-3}{{\em Phys. Lett.}
  {\bfseries B477} (2000) 147--155},
\href{http://arxiv.org/abs/hep-ph/9911418}{{\ttfamily arXiv:hep-ph/9911418
  [hep-ph]}}.

\bibitem{Branco:2010tx}
G.~C. Branco, D.~Emmanuel-Costa, and C.~Simoes, ``{Nearest-Neighbour
  Interaction from an Abelian Symmetry and Deviations from Hermiticity}'',
  \href{http://dx.doi.org/10.1016/j.physletb.2010.05.009}{{\em Phys. Lett.}
  {\bfseries B690} (2010) 62--67},
\href{http://arxiv.org/abs/1001.5065}{{\ttfamily arXiv:1001.5065 [hep-ph]}}.

\bibitem{Emmanuel-Costa:2016gdp}
D.~Emmanuel-Costa and R.~Gonzalez~Felipe, ``{More about unphysical zeroes in
  quark mass matrices}'',
  \href{http://dx.doi.org/10.1016/j.physletb.2016.11.019}{{\em Phys. Lett.}
  {\bfseries B764} (2017) 150--156},
\href{http://arxiv.org/abs/1609.09491}{{\ttfamily arXiv:1609.09491 [hep-ph]}}.

\bibitem{Ishimori:2010au}
H.~Ishimori, T.~Kobayashi, H.~Ohki, Y.~Shimizu, H.~Okada, and M.~Tanimoto,
  ``{Non-Abelian Discrete Symmetries in Particle Physics}'',
  \href{http://dx.doi.org/10.1143/PTPS.183.1}{{\em Prog. Theor. Phys. Suppl.}
  {\bfseries 183} (2010) 1--163},
\href{http://arxiv.org/abs/1003.3552}{{\ttfamily arXiv:1003.3552 [hep-th]}}.

\bibitem{Rasin:1998je}
A.~Rasin, ``{Hierarchical quark mass matrices}'',
  \href{http://dx.doi.org/10.1103/PhysRevD.58.096012}{{\em Phys. Rev.}
  {\bfseries D58} (1998) 096012},
\href{http://arxiv.org/abs/hep-ph/9802356}{{\ttfamily arXiv:hep-ph/9802356
  [hep-ph]}}.

\bibitem{Froggatt:1978nt}
C.~D. Froggatt and H.~B. Nielsen, ``{Hierarchy of Quark Masses, Cabibbo Angles
  and CP Violation}'',
\href{http://dx.doi.org/10.1016/0550-3213(79)90316-X}{{\em Nucl. Phys.}
  {\bfseries B147} (1979) 277--298}.

\bibitem{ArkaniHamed:1999dc}
N.~Arkani-Hamed and M.~Schmaltz, ``{Hierarchies without symmetries from extra
  dimensions}'', \href{http://dx.doi.org/10.1103/PhysRevD.61.033005}{{\em Phys.
  Rev.} {\bfseries D61} (2000) 033005},
\href{http://arxiv.org/abs/hep-ph/9903417}{{\ttfamily arXiv:hep-ph/9903417
  [hep-ph]}}.

\bibitem{Couture:2009it}
G.~Couture, C.~Hamzaoui, S.~S.~Y. Lu, and M.~Toharia, ``{Patterns in the
  Fermion Mixing Matrix, a bottom-up approach}'',
  \href{http://dx.doi.org/10.1103/PhysRevD.81.033010}{{\em Phys. Rev.}
  {\bfseries D81} (2010) 033010},
\href{http://arxiv.org/abs/0910.3132}{{\ttfamily arXiv:0910.3132 [hep-ph]}}.

\bibitem{Branco:2011aa}
G.~C. Branco and J.~I. Silva-Marcos, ``{Invariants, Alignment and the Pattern
  of Fermion Masses and Mixing}'',
  \href{http://dx.doi.org/10.1016/j.physletb.2012.07.064}{{\em Phys. Lett.}
  {\bfseries B715} (2012) 315--321},
\href{http://arxiv.org/abs/1112.1631}{{\ttfamily arXiv:1112.1631 [hep-ph]}}.

\bibitem{Cheng:1987rs}
T.~P. Cheng and M.~Sher, ``{Mass Matrix Ansatz and Flavor Nonconservation in
  Models with Multiple Higgs Doublets}'',
\href{http://dx.doi.org/10.1103/PhysRevD.35.3484}{{\em Phys. Rev.} {\bfseries
  D35} (1987) 3484}.

\bibitem{deMedeirosVarzielas:2017sdv}
I.~de~Medeiros~Varzielas, G.~G. Ross, and J.~Talbert, ``{A Unified Model of
  Quarks and Leptons with a Universal Texture Zero}'',
\href{http://arxiv.org/abs/1710.01741}{{\ttfamily arXiv:1710.01741 [hep-ph]}}.

\bibitem{Branco:1992ba}
G.~C. Branco and F.~Mota, ``{Heavy top, Fritzsch ansatz and radiative
  corrections}'',
\href{http://dx.doi.org/10.1016/0370-2693(92)90781-X}{{\em Phys. Lett.}
  {\bfseries B280} (1992) 109--112}.

\bibitem{Weinberg:1979sa}
S.~Weinberg, ``{Baryon and Lepton Nonconserving Processes}'',
\href{http://dx.doi.org/10.1103/PhysRevLett.43.1566}{{\em Phys.Rev.Lett.}
  {\bfseries 43} (1979) 1566--1570}.

\bibitem{Branco:2007nn}
G.~C. Branco, D.~Emmanuel-Costa, R.~Gonzalez~Felipe, and H.~Serodio, ``{Weak
  Basis Transformations and Texture Zeros in the Leptonic Sector}'',
  \href{http://dx.doi.org/10.1016/j.physletb.2008.10.059}{{\em Phys. Lett.}
  {\bfseries B670} (2009) 340--349},
\href{http://arxiv.org/abs/0711.1613}{{\ttfamily arXiv:0711.1613 [hep-ph]}}.

\bibitem{Altarelli:2005yp}
G.~Altarelli and F.~Feruglio, ``{Tri-bimaximal neutrino mixing from discrete
  symmetry in extra dimensions}'',
  \href{http://dx.doi.org/10.1016/j.nuclphysb.2005.05.005}{{\em Nucl. Phys.}
  {\bfseries B720} (2005) 64--88},
\href{http://arxiv.org/abs/hep-ph/0504165}{{\ttfamily arXiv:hep-ph/0504165
  [hep-ph]}}.

\bibitem{Altarelli:2005yx}
G.~Altarelli and F.~Feruglio, ``{Tri-bimaximal neutrino mixing, A(4) and the
  modular symmetry}'',
  \href{http://dx.doi.org/10.1016/j.nuclphysb.2006.02.015}{{\em Nucl. Phys.}
  {\bfseries B741} (2006) 215--235},
\href{http://arxiv.org/abs/hep-ph/0512103}{{\ttfamily arXiv:hep-ph/0512103
  [hep-ph]}}.

\bibitem{Harrison:2002er}
P.~F. Harrison, D.~H. Perkins, and W.~G. Scott, ``{Tri-bimaximal mixing and the
  neutrino oscillation data}'',
  \href{http://dx.doi.org/10.1016/S0370-2693(02)01336-9}{{\em Phys. Lett.}
  {\bfseries B530} (2002) 167},
\href{http://arxiv.org/abs/hep-ph/0202074}{{\ttfamily arXiv:hep-ph/0202074
  [hep-ph]}}.

\bibitem{Ahn:2012nd}
{\bfseries RENO} , J.~K. Ahn {\em et~al.}, ``{Observation of Reactor Electron
  Antineutrino Disappearance in the RENO Experiment}'',
  \href{http://dx.doi.org/10.1103/PhysRevLett.108.191802}{{\em Phys. Rev.
  Lett.} {\bfseries 108} (2012) 191802},
\href{http://arxiv.org/abs/1204.0626}{{\ttfamily arXiv:1204.0626 [hep-ex]}}.

\bibitem{An:2012eh}
{\bfseries Daya Bay} , F.~P. An {\em et~al.}, ``{Observation of
  electron-antineutrino disappearance at Daya Bay}'',
  \href{http://dx.doi.org/10.1103/PhysRevLett.108.171803}{{\em Phys. Rev.
  Lett.} {\bfseries 108} (2012) 171803},
\href{http://arxiv.org/abs/1203.1669}{{\ttfamily arXiv:1203.1669 [hep-ex]}}.

\bibitem{Abe:2013hdq}
{\bfseries T2K} , K.~Abe {\em et~al.}, ``{Observation of Electron Neutrino
  Appearance in a Muon Neutrino Beam}'',
  \href{http://dx.doi.org/10.1103/PhysRevLett.112.061802}{{\em Phys. Rev.
  Lett.} {\bfseries 112} (2014) 061802},
\href{http://arxiv.org/abs/1311.4750}{{\ttfamily arXiv:1311.4750 [hep-ex]}}.

\bibitem{Esteban:2016qun}
I.~Esteban, M.~C. Gonzalez-Garcia, M.~Maltoni, I.~Martinez-Soler, and
  T.~Schwetz, ``{Updated fit to three neutrino mixing: exploring the
  accelerator-reactor complementarity}'',
  \href{http://dx.doi.org/10.1007/JHEP01(2017)087}{{\em JHEP} {\bfseries 01}
  (2017) 087}, \href{http://arxiv.org/abs/1611.01514}{{\ttfamily
  arXiv:1611.01514 [hep-ph]}}, NuFIT 3.2 (2018), \texttt{www.nu-fit.org}.

\bibitem{deSalas:2017kay}
P.~F. de~Salas, D.~V. Forero, C.~A. Ternes, M.~Tortola, and J.~W.~F. Valle,
  ``{Status of neutrino oscillations 2017}'',
\href{http://arxiv.org/abs/1708.01186}{{\ttfamily arXiv:1708.01186 [hep-ph]}}.

\end{thebibliography}\endgroup

\end{document}